\RequirePackage{fix-cm}
\documentclass[twocolumn]{svjour3}          
\smartqed  
\usepackage{graphicx}
\usepackage{newtxtext,newtxmath}
\journalname{Computational Geosciences}

\usepackage{amsmath}
\usepackage{amsfonts}
\usepackage{float}
\usepackage{booktabs}
\usepackage{natbib}
\usepackage[colorlinks=true,allcolors=blue]{hyperref}
\usepackage{float}
\usepackage{multirow}
\usepackage[dvipsnames]{xcolor}

\DeclareSymbolFont{letters}{OML}{cmm}{m}{it}
\DeclareMathAlphabet{\mathcal}{OMS}{cmsy}{m}{n}

\newcommand{\edit}[1]{{\color{black}#1}}
\newcommand{\editone}[1]{{\color{black}#1}}
\newcommand{\edittwo}[1]{{\color{black}#1}}
\newcommand{\editfour}[1]{{\color{black}#1}}

\begin{document}

\title{Compositional Effects in Thermal, Compositional and Reactive Simulation}


\author{Matthias A. Cremon         \and
        Margot G. Gerritsen}


\institute{Matthias A. Cremon \at
              \email{mcremon@stanford.edu}           
           \and
           Margot G. Gerritsen \at
              \email{margot.gerritsen@stanford.edu}
}

\date{Received: date / Accepted: date}

\maketitle

\begin{abstract}
This work studies the influence of several compositional effects on thermal and reactive processes.

First, \edit{the impact of using a fully compositional model in the context of thermal simulations is considered.} Detailed phase behavior models rely on compositional descriptions of the oil using up to tens of components. Lumping a large number of components into a smaller number of pseudo-components in order to reduce the computational cost is standard practice for thermal simulations. \edit{Lumping schemes are typically calibrated using experimental data, in order to achieve a good approximation of the phase behavior of the initial oil. Due to the evaporation and condensation of components under thermal stimulation, the oil composition will widely vary in time and space. This works illustrates that even if the lumped schemes were able to capture the phase behavior of the initial oil, the lack of resolution can lead to modeling artefacts and/or fail to capture the relevant displacement processes.} 

Then, the effects of different compositional interpretations of the lumped pseudo-species appearing in typical reaction schemes are investigated.
A constant, mass-based fraction of the oil is allowed to react, and the reactive components are modified. Due to molecular weight effects, the reaction rate is larger when light and medium components are allowed to react. \edit{Those effects result in an increased displacement, leading to faster fronts, larger oil banks and more pressurization of the system.}
\end{abstract}

\section{Introduction}

Thermal recovery of heavy oil uses heat, either injected in steam drive or Steam Assisted Gravity Drainage (SAGD) processes, or generated in-situ via exothermic (combustion) reactions \citep{Prats82}. The higher temperature lowers the viscosity of the oil and allows it to be displaced. When crude oil is subjected to higher temperatures, light and intermediate components vaporize, move downstream in the vapor phase and then condense back to the liquid phase. This behavior is a key part of the thermal recovery processes, and cannot be properly captured without using a detailed compositional description of the oil. In compositional simulation, to reduce the computational cost, it is standard practice to lump pure components into a smaller number of pseudo-components when running the flow and transport problem \citep{Kay36,Montel84,Nowley91,Leibovici93,Jessen07,Rastegar09,Pedersen14}. For thermal simulation, the computational work is even greater due to the addition of the temperature variable and the added coupling between the mass and the energy equation.

This work first investigates the impact of the number of lumped pseudo-components on the thermal recovery of heavy oil. \citet{Lapene11} identified the need for a compositional description of heavy oil in the context of static Ramped Temperature Oxidation (RTO) experiments. That work is extended to displacement processes to study the impact of the lumping strategy on 1D hot nitrogen injection and in-situ combustion cases\edit{, using lumping schemes calibrated with experimental data available for the initial oil}. To estimate the efficiency of recovery methods, capturing the displacement processes is crucial. A combustion reaction scheme from \citet{Dechelette06} is used to study the lumping effects in the presence of exothermic reactions. The non-linearity and tight coupling of the problem leads to a loss in accuracy when using aggressive lumping of hydrocarbon pseudo-components. \edit{The implementation of the model into Stanford's Automatic Differentiation General Purpose Research Simulator (AD-GPRS, \citet{adgprs}) makes it possible to run simulations using the most detailed composition of the oil samples, which in this work results in 60 unknowns per cell when running thermal, compositional and reactive cases. To the best of our knowledge, this work is the first attempt to use such detailed compositional descriptions in dynamic simulations.}


\editone{The general goal of this work is to identify possible issues of using a detailed compositional model in thermal and reactive simulations. Reaction schemes published in the literature typically do not use more than a couple oil species \citep{Crookston79, Coats80b, Dechelette06, Cinar11}, and it is not straightforward to map the components used in the compositional model to the species used in the reaction schemes. The second part of this manuscript presents the first study using several different compositional mappings for an existing reaction scheme.} Although the scheme from \citet{Dechelette06} uses only a single pseudo-species, Oil, different components are allowed to be reactants or not it is observed that when medium components can react, the front speeds are larger, the oil banks bigger and the pressure rises.

The paper is structured as follows. Section~\ref{sec::mathmod} lists the governing equations and local constraints used in the simulations. Section~\ref{sec::lumpingResults} presents the oil samples and the lumping numerical results obtained using three test cases. Section~\ref{sec::reactionResults} shows the results of changing the compositional description of the chemical reaction schemes. Finally, a conclusion and ideas for future work are given in Section~\ref{sec::conclusion}.

\section{Mathematical Model}
\label{sec::mathmod}

This section briefly describes the Partial Differential Equations (PDEs) used to model the compositional and thermal displacement of fluids in porous media. A detailed description of the governing equations for compositional reservoir simulation can be found in \citet{Cao02}, among many others, and the thermal and reactive formulation is described in \citet{Coats80b} and \citet{Cremon20thesis}.

\subsection{Conservation Equations}

A compositional formulation is used, for which the mass is conserved for each component across all phases (oil, water, gas). It considers $n_c$ fluid components in $n_p$ fluid phases in the general case, leading to
\begin{equation} \label{eq::mass}
 \dfrac{\partial}{\partial t}\left(\phi\sum\limits_{p\,=\,1}^{n_p}x_{cp}\rho_pS_p\right) + \nabla \cdot \left(\sum\limits_{p\,=\,1}^{n_p}x_{cp}\rho_p\textbf{u}_p\right) + q_c^\textrm{w} + q_c^\textrm{r} = 0,
\end{equation}
for $c = 1,\dots,n_c$ and where $p$ and $c$ are the phase and component indices, $\phi$ is the porosity, $x_{cp}$ is the mole fraction of component $c$ in phase $p$, $\rho_p$, $S_p$ and $\textbf{u}_p$ are the molar density, saturation and velocity of phase $p$, $q_c^\textrm{w}$ is the source term from wells and $q_c^\textrm{r}$ the source term from reactions.

The $n_s$ solid components are considered immobile reaction products and obey a simpler mass conservation equation
\begin{equation} \label{eq::mass_solid}
 \dfrac{\partial}{\partial t}\left(\phi c_s\right) + q_s^\textrm{r} = 0, \qquad s = 1,\dots,n_s,
\end{equation}
where $c_s$ is the molar concentration and $q_s^\textrm{r}$ the source term from reactions.

In the thermal formulation, the energy is conserved according to
\begin{align} \label{eq::energy}
\begin{split}
\dfrac{\partial}{\partial t}\left(\phi\sum\limits_{p\,=\,1}^{n_p}U_{p}\rho_pS_p+(1-\phi)\tilde U_R\right) + \nabla \cdot \left(\sum\limits_{p\,=\,1}^{n_p}H_{p}\rho_p\textbf{u}_p\right) \\ - \nabla \cdot \left(\kappa\nabla T\right) + q^\textrm{w} + q^\textrm{r} + q^\textrm{hl} + q^\textrm{hr} = 0,
\end{split}
\end{align}
where $p$ and $c$ are the phase and component indices, T is the temperature, $\tilde U_R$ is the rock volumetric internal energy, $\kappa$ is the thermal conductivity, $U_{p}$ and $H_{p}$ are the internal energy, enthalpy of phase $p$, $q^\textrm{w}$, $q^\textrm{r}$, $q^\textrm{hl}$ and $q^\textrm{hr}$ are the source terms from wells, reactions, heat losses and heater, respectively. Thermal diffusion (also called conduction) cannot be neglected due to the large heat conductivity of the rock matrix \citep{Prats82}. 

\subsection{Local Constraints}

The natural variables \citep{Coats80,Cao02,Voskov12b} formulation used in this work needs additional local (cell-based) constraints to obtain a well-posed system. In each phase, the sum of all molar fractions shall be equal to 1, giving $n_p$ equations
\begin{equation} \label{eq::phase}
\sum\limits_{c\,=\,1}^{n_c} x_{cp} = 1,\qquad p = 1,\dots,n_p.
\end{equation}
For each component $c$, the thermodynamic equilibrium condition states that the chemical potential $\mu_c$ in all present phases must be equal, which reads
\begin{equation} \label{eq::chempot}
\mu_{cj} = \mu_{ck},\ \forall j\neq k,\qquad c = 1,\dots,n_c,
\end{equation}
where $c$ is the component index, $j$ and $k$ are phases indices, and $\mu_{cp}$ is the chemical potential of component $c$ in phase $p$. When using an Equation of State (EoS) model to compute phase properties, it is convenient to recast that equation in terms of fugacity \citep{Michelsen82a} and obtain
\begin{equation} \label{eq::fug}
f_{cj} = f_{ck},\ \forall j\neq k,\qquad c = 1,\dots,n_c,
\end{equation}
where $f_{cp}$ is the fugacity of component $c$ in phase $p$. Unlike chemical potential, fugacity can be readily computed from Equation of State (EoS) parameters. All of our simulations use the Peng-Robinson EoS \citep{Peng76}.

\subsection{Phase Behavior}

A free-water (FW) flash initially presented in \citet{Lapene10} is used for the phase behavior calculations. For heavy oil recovery using thermal methods, the solubility of oil in water is negligible:
\begin{subequations}
\begin{align}
z_i & = \  x_iO + \ y_iV, \hspace{1cm} i = 1,\dots,n_c,\ i \neq w  \,, \label{CPTR_eq::FW1} \\
z_w & = x_wO + y_wV + W  \,, \label{CPTR_eq::FW2}
\end{align}
\end{subequations}
with $O$, $V$ and $W$ the oil, vapor and water phase molar fractions and $z_i$ the overall mole fraction of component $i$.

\section{Effects of Lumping on the Displacement Process}
\label{sec::lumpingResults}

This section presents numerical results for two oils and three test cases: Hot Nitrogen Injection (HNI, no reactions), Hot Air Injection (HAI) and Cold Air Injection with Heaters (CAIH).

\subsection{Oil Samples}

\begin{figure}[t!]
    \centering
    \includegraphics[width=.485\textwidth]{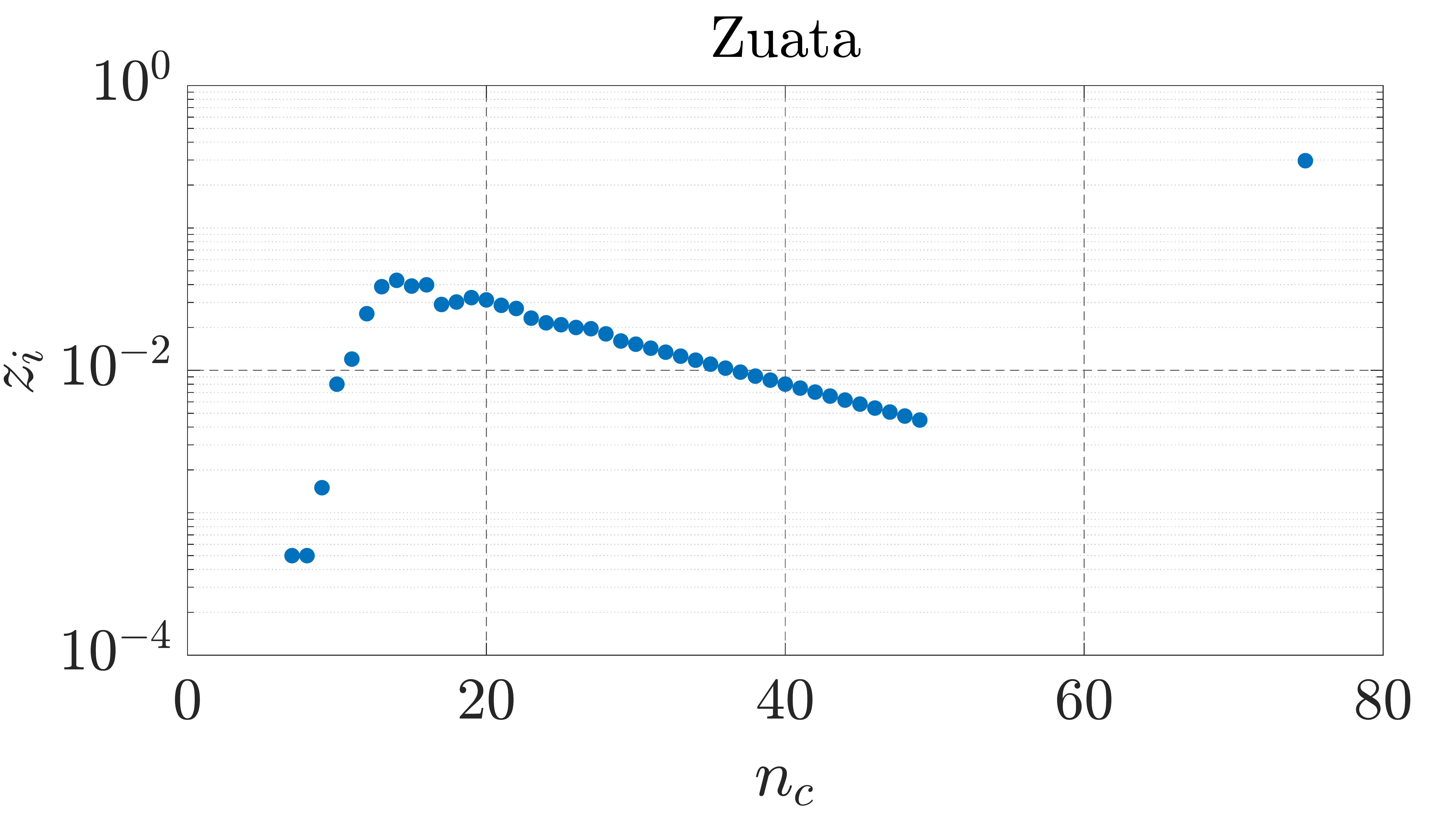}
    
    \includegraphics[width=.485\textwidth]{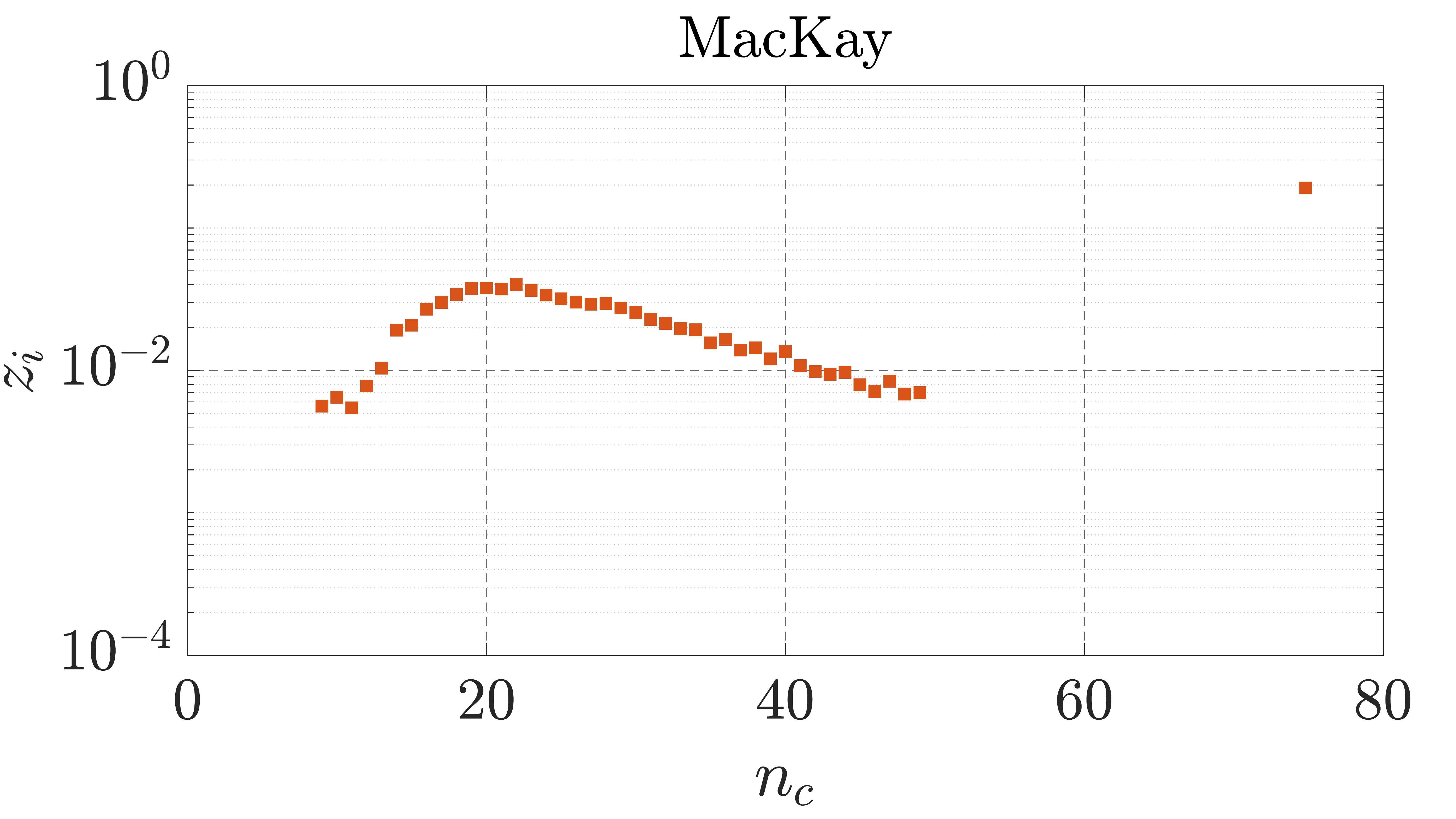}
    \caption{Molar fractions of Zuata (top) and MacKay (bottom) oil samples, as a function of carbon number (hydrocarbons only).}
    \label{fig::compositions}
\end{figure}

The first sample is from the Zuata field, in the Orinoco Belt region of Venezuela, described at length in \citet{Lapene10b}. It is an extra heavy oil, with a density of 8.5$^\circ$API and a viscosity of 10,000 cP at reservoir conditions (48$^\circ$C and 50 bars). The second sample is from MacKay, Alberta in Canada, described in \citet{Nourozieh13}. It is slightly lighter, with a density of 12.9$^\circ$API but shows a higher viscosity than Zuata, especially at low temperatures, which is why it is labeled as a bitumen. The global molar fractions of both oils is plotted in Fig.~\ref{fig::compositions}. They are considered dead at laboratory conditions, with the first hydrocarbon components present being C$_7$ and C$_9$, respectively. \editone{Using two samples ensures that the observations are not oil-dependant.}


These oils are virtually impossible to displace using non-thermal recovery methods, due to their viscosity below 50$^\circ$C being largely above 10,000 cP. Depending on the depth of the reservoirs, steam injection can be used, but these two oils are interesting candidates for in-situ combustion. Detailed compositions for both those oils are available, containing 52 hydrocarbon components. \edit{For more details about the properties of the oils, the reader is referred to \citet{Cremon20thesis}}. Those detailed compositions, due to the increased size of the linear systems (especially with natural variables), are impractical to run simulations. Using AD-GPRS \citep{adgprs}, a 1D detailed run (400 cells) takes more than a day \edit{using a CPU-parallel (OpenMP) implementation on a standard workstation (4-core x64 Intel Core i7-4790 3.6 Ghz, 16 GB of DDR3-1600 RAM)}.
The compositions are typically lumped into less than 10 hydrocarbon pseudo-components to keep the computational cost reasonable. The results section expands on the runtime considerations.

An 8-component lumping from \citet{Lapene10b} is used, as well as a 4-component lumping. A 2-component lumping led to severely inaccurate results, with the initial flash with two components yields 20\% more oil than other lumpings. All lumpings are given in Table \ref{tab::lumpings}. The 4-component lumped results are compared with the detailed composition for two test cases in the next section. Each oil is mixed with 90\% water and 1\% nitrogen (by mole), so that a realistic oil, gas and water saturation for a combustion tube experiment is obtained under our initial conditions.


\begin{figure*}[t!]
    \centering
    \includegraphics[width=\textwidth]{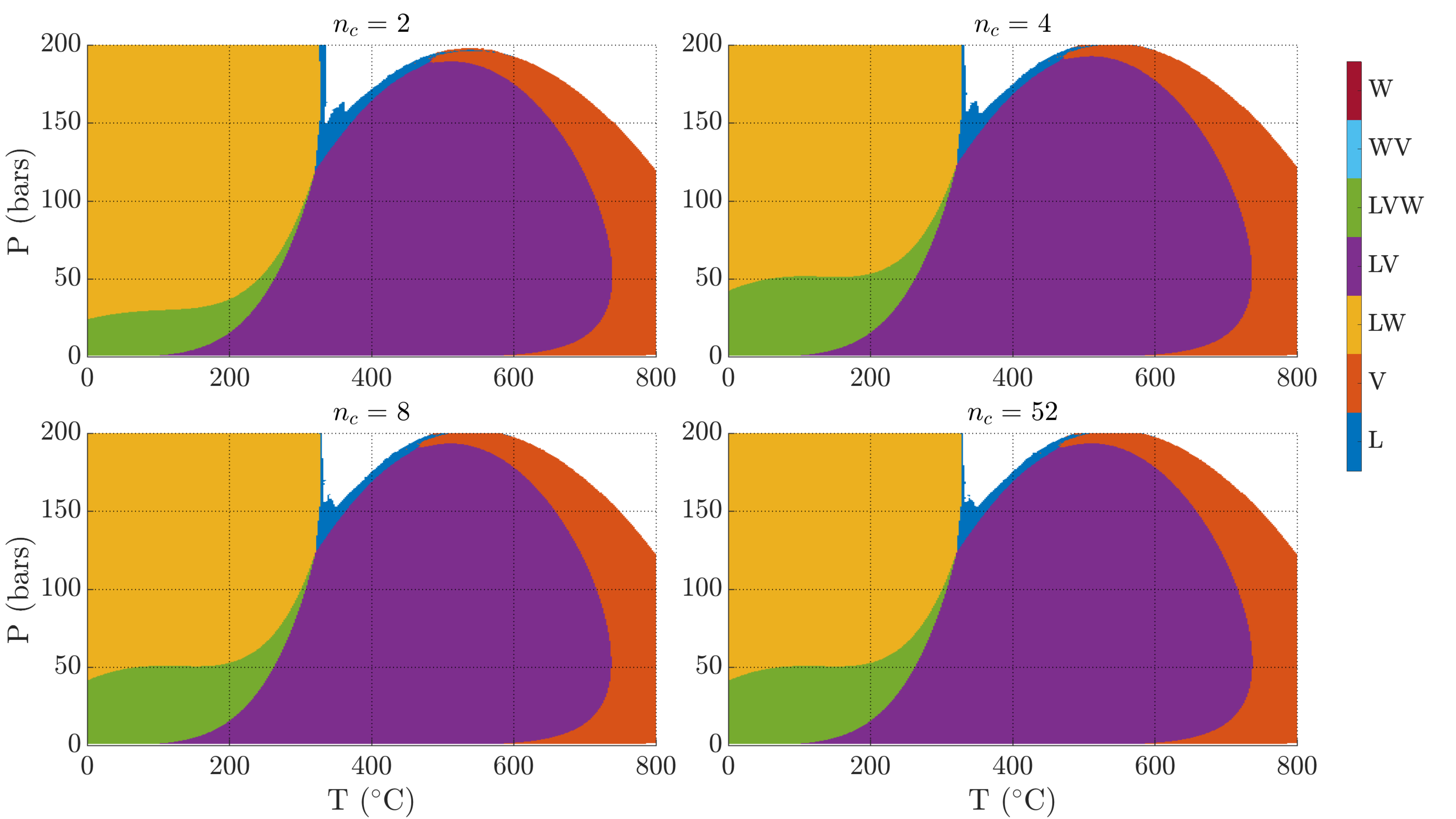}
    \caption{\edittwo{Phase diagrams for the Dead Zuata oil, using 2 (top left), 4 (top right), 8 (bottom left) and 52 (bottom right) components. Key: water phase (W), liquid oil phase (L) and vapor phase (V).}}
    \label{fig::DeadZuataEnvelopes}
\end{figure*}

\edittwo{Tuning properties of the hydrocarbon pseudo-components is standard practice. Following the methodology of \citet{Pedersen14}, the molecular weight of the heavy fraction is modified in order to match the phase behavior obtained through experiments. In the case of the Zuata oil, properties of the C$_{50+}$ fraction were adjusted to yield a phase behavior consistent with the experimental data -- for the initial oil \citep{Lapene10b}. The phase diagrams obtained using all lumpings for Zuata oil is shown in Fig. \ref{fig::DeadZuataEnvelopes}. Both the 4- and 8-component lumpings yield phase diagrams virtually identical to the 52-component results for the initial oil. The 2-component lumping does not capture the phase behavior as well, especially the three-phase region (below 50 bars and below 300$^\circ$C), which includes our initial conditions.}


\edit{The remainder of this section illustrates that matching the initial oil phase behavior is not a sufficient condition to capture the dynamic displacement process for thermal recovery cases. The performance deteriorates quickly at higher temperatures and large errors, of $\mathcal{O}(1)$, are observed on the phase molar fractions due to the lack of resolution within the pseudo-components \citep{Cremon21a}.
}

\begin{table}[hb!]
    \centering
    \caption{List of pseudo-components for the different lumpings.}
    \begin{tabular}{ccc}
    \toprule
        2-component & 4-component & 8-component \\
        \midrule
        \multirow{8.5}{*}{C$_1$-C$_{49}$} & C$_1$ & C$_1$ \\
         \cmidrule{2-3}
         & \multirow{2.5}{*}{C$_{2}$-C$_{16}$} & C$_{2}$-C$_{11}$ \\
         \cmidrule{3-3}
         & & C$_{12}$-C$_{16}$ \\
         \cmidrule{2-3}
         & \multirow{5.5}{*}{C$_{17}$-C$_{49}$} & C$_{17}$-C$_{21}$ \\
         \cmidrule{3-3}
         & & C$_{22}$-C$_{26}$ \\
         \cmidrule{3-3}
         & & C$_{27}$-C$_{35}$ \\
         \cmidrule{3-3}
         & & C$_{36}$-C$_{49}$ \\
        \midrule
        C$_{50+}$ & C$_{50+}$ & C$_{50+}$ \\
        \bottomrule
    \end{tabular}
    \label{tab::lumpings}
\end{table}

\subsection{Hot Nitrogen Injection (HNI)}

Our first test case, HNI, is a thermal-compositional hot nitrogen injection problem, without reactions. Pure nitrogen is injected at 600$^\circ$C into a mixture of oil, water and nitrogen.


\begin{figure*}[t!]
    \centering
    \includegraphics[width=\textwidth]{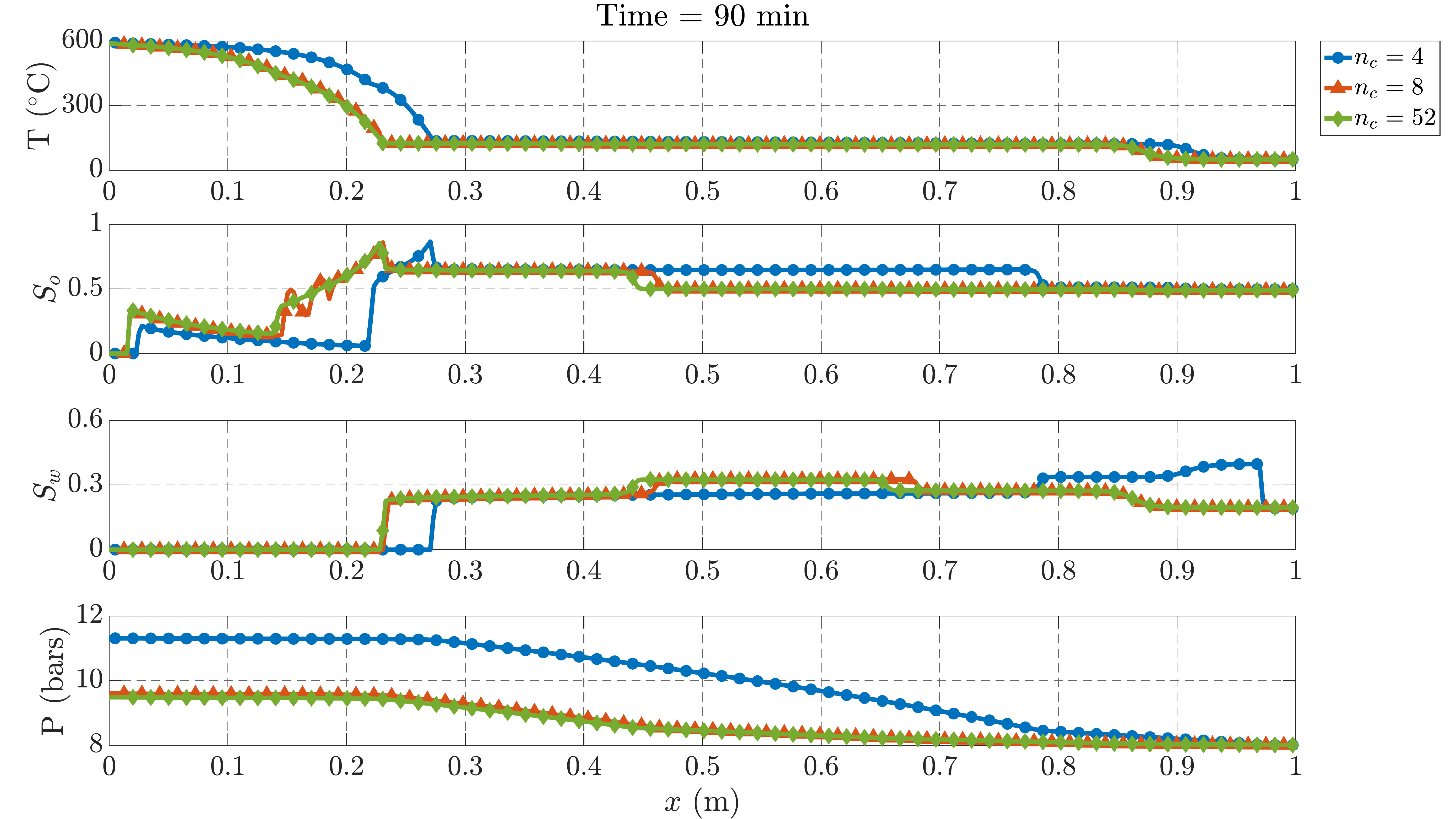}
    \caption{Hot nitrogen injection results for Zuata oil. Temperature (top), oil saturation (middle top), water saturation (middle bottom) and pressure (bottom) using 4 components (blue circles), 8 components (orange triangles) and 52 components (green diamonds).}
    \label{fig::DeadZuata_Inj600CNoReac}
\end{figure*}

\begin{figure*}[t!]
    \centering
    \includegraphics[width=\textwidth]{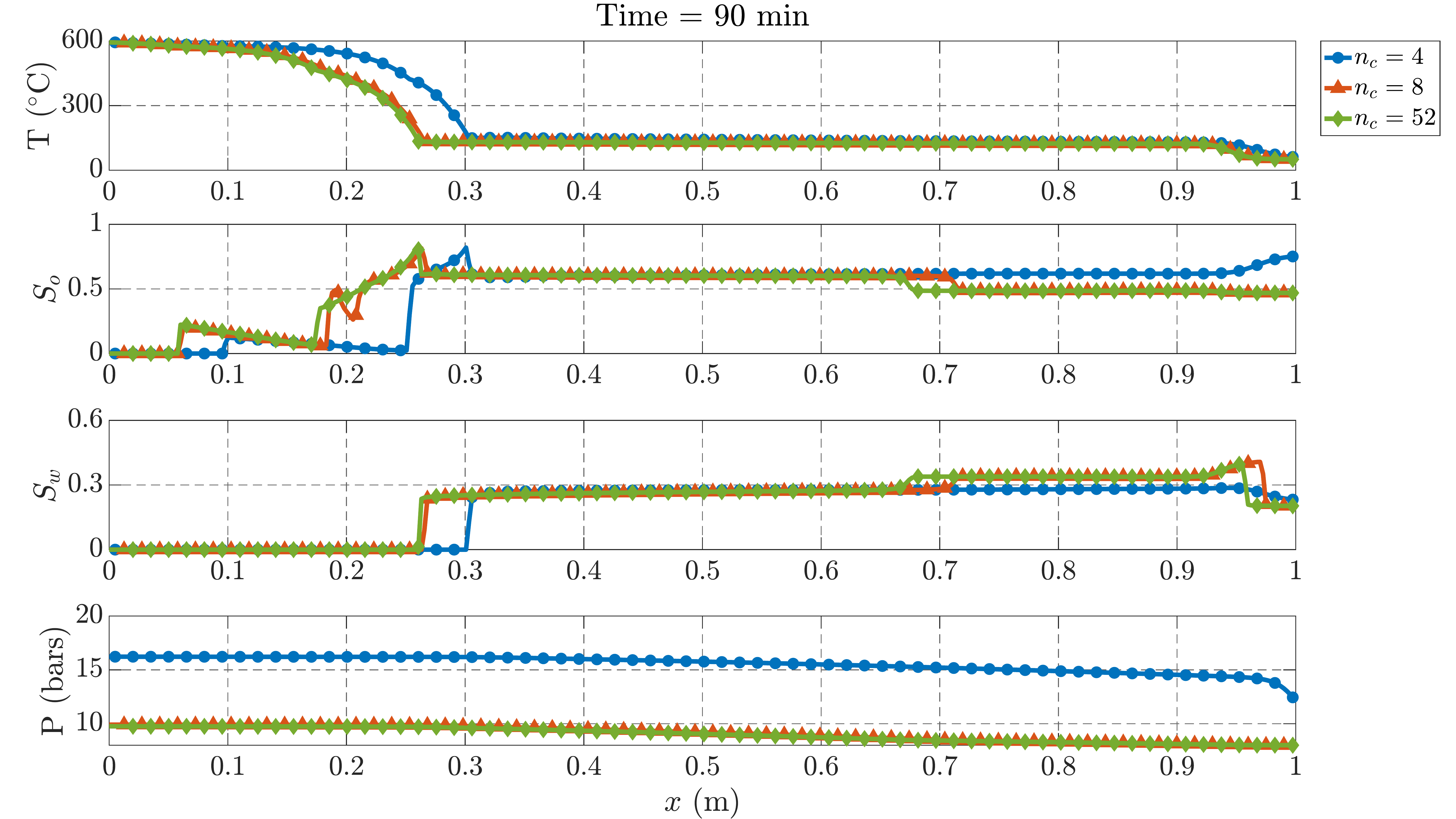}
    \caption{Hot nitrogen injection results for MacKay oil. Temperature (top), oil saturation (middle top), water saturation (middle bottom) and pressure (bottom) using 4 components (blue circles), 8 components (orange triangles) and 52 components (green diamonds).}
    \label{fig::MacKayBitumen_Inj600CNoReac}
\end{figure*}

Figure~\ref{fig::DeadZuata_Inj600CNoReac} and \ref{fig::MacKayBitumen_Inj600CNoReac} show the temperature, oil saturation, water saturation and pressure profiles for the Zuata oil and the MacKay oil after 90 min of nitrogen injection, respectively. Injecting hot nitrogen (or cold nitrogen paired with electrical heaters) is the standard way of pre-heating a combustion tube \citep{Yoo19}. The flash at initial conditions leads to less than 2\% 1-norm error on the oil saturation with the 8-component lumping for both oils, and less than 5\% error with the 4-component lumping. A detailed description of the parameters is given in Table \ref{tab::hot-air}.

\begin{table}[b!]
    \centering
    \caption{Simulation parameters for hot nitrogen injection cases, in S.I. and laboratory units.}
    {
    \begin{tabular}{llrlrl}
    \toprule
    Property & Symbol & S.I. & & Lab &  \\
    \midrule
    Domain Size & $L$ & 1 & m & 1 & m \\
    Porosity & $\phi$ & 0.36 & -- & 0.36 & -- \\
    Permeability & $k$ & 10$^{-12}$ & m$^2$ & 10 & D \\
    Injection Rate & $q$ & 4.32 & m$^3$/day & 3 & L/min \\
    Injection Temp. & $T_\textrm{inj}$ & 873.15 & $^\circ$K & 600 & $^\circ$C \\
    Initial Temp. & $T_\textrm{init}$ & 323.15 & $^\circ$K & 50 & $^\circ$C \\
    Initial Pressure & $P_\textrm{init}$ & 8 & bar & 116 & psi \\
    \bottomrule
    \end{tabular}
    }
    \label{tab::hot-air}
\end{table}

First and foremost, the 4-component lumping results are off on most quantities for both oils. The pressure is overestimated by about 20\% for Zuata and 60\% for MacKay, and all the fronts (steam, oil banks, water banks and evaporation, trailing oil evaporation) show inaccurate speeds compared to the detailed 52 components case. The 8-component lumping fares a lot better and captures most of the crucial features of the displacement. By accurately estimating the pressure, the front speeds are close to the 52 components case. The only slight difference is seen with the oil bank front, around 0.45m. Since the medium components are lumped for the 8-component case, a different front speed is observed. Due to compositional effects in the \editone{density} calculations, the two-phase region (0.05 to 0.22) can show \editone{numerical artefacts in the saturation  around the smooth 52-component oil saturation curve, due to components molar fractions moving as bell curves in the domain and impacting the gas density calculations.}

The differences between the 4- and 8-component lumpings are in the treatment of the intermediate components. In both cases, C$_1$ and C$_{50+}$ are retained as pure components, but the for the 8-component case, light oil (C$_{2}$-C$_{16}$) and medium oil (C$_{17}$-C$_{49}$) are further decomposed, into two and four components, respectively. This increased resolution allows the sequential, temperature dependent evaporation/condensation process to occur in a smoother way, and retains enough pseudo-components to closely match the 52-component case.

\subsection{Hot Air Injection (HAI)}

\begin{table}[b!]
    \centering
    \caption{Reaction parameters for the in-situ combustion test case: frequency factor (A), activation energy (E$_a$) and enthalpy of reaction ($h_r$)}
    {
    \begin{tabular}{lcrrr}
    \toprule
     &  & A & E$_a$ & $h_r$ \\
    Reaction & \# & (J/mol) & (min$^{-1}$.kPa$^{-1}$) & (MJ) \\
    \midrule
    FD/LTO & 1 & 25,000 & 1  & 1,600 \\
    HTO$_1$ & 2 & 67,000 & 250  & 12,800 \\
    HTO$_2$ & 3 & 87,000 & 220  & 4,850 \\
    \bottomrule
    \end{tabular}
    }
    \label{tab::dech}
\end{table}

The next case considered, HAI, includes chemical reactions and simulates an in-situ combustion process. The reaction scheme is from \citet{Dechelette06}, with a combined fuel deposition and low-temperature oxidation (FD/LTO) reaction and two high-temperature oxidation (HTO) reactions burning solid fuel:
\begin{subequations}
\begin{align}
 \textrm{Oil} + \textrm{O$_2$} & \rightarrow \textrm{Coke$_1$}, \label{eq::FD}\\
 \textrm{Coke$_1$} + \textrm{O$_2$} & \rightarrow \textrm{Coke$_2$} + \textrm{CO$_2$} + \textrm{CO} + \textrm{H$_2$O}, \\
 \textrm{Coke$_2$} + \textrm{O$_2$} & \rightarrow \textrm{CO$_2$} + \textrm{CO} + \textrm{H$_2$O},
\end{align}
\end{subequations}
where the stoichiometry factors are omitted for simplicity. All equations consume oxygen and release heat (exothermic). The reaction parameters are given in Tab.~\ref{tab::dech}. Air is injected at 400$^\circ$C to reach ignition; every other parameter is the same as in the hot nitrogen injection case (see Tab.~\ref{tab::hot-air}).


\begin{figure*}[t!]
    \centering
    \includegraphics[width=\textwidth]{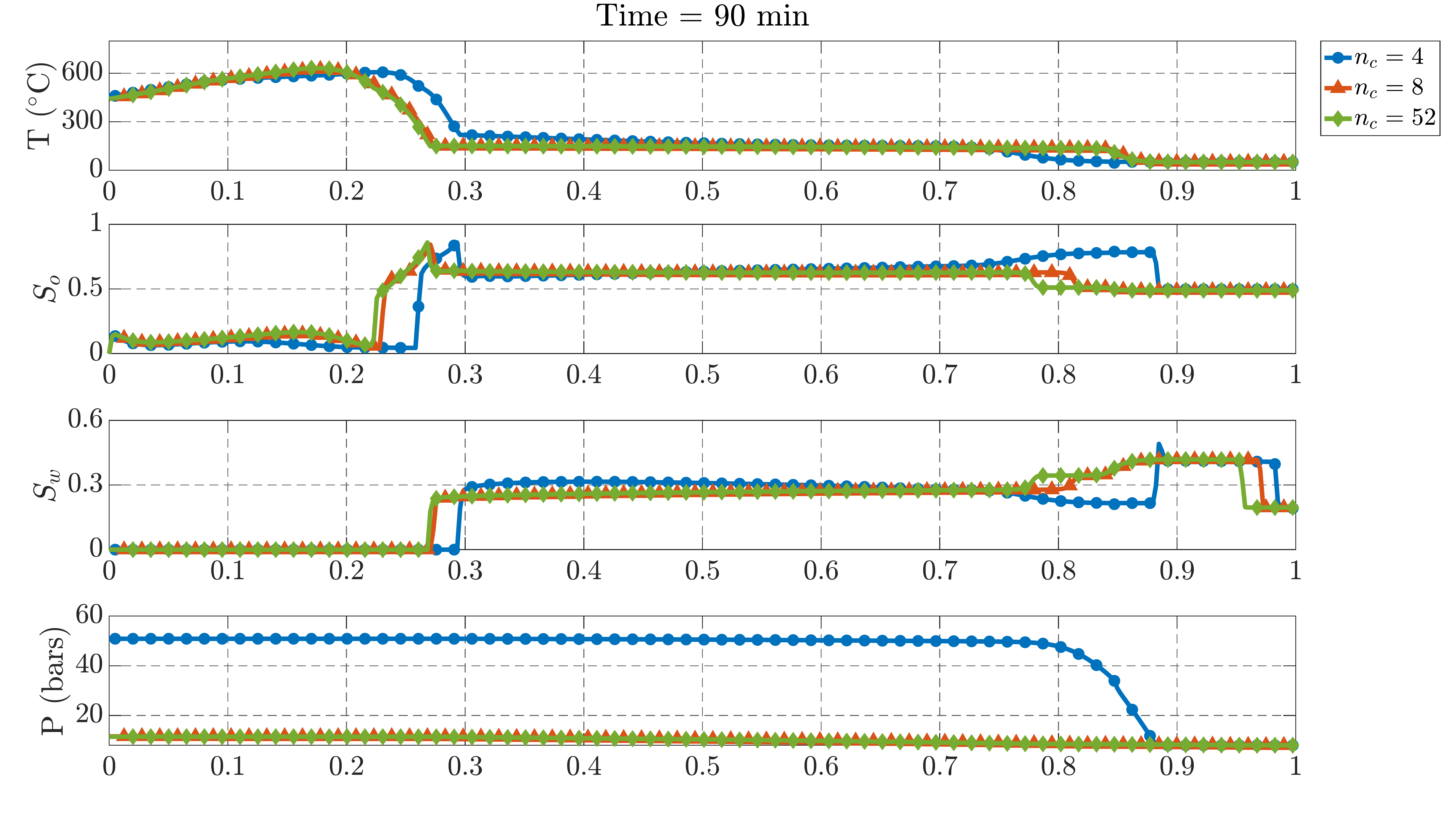}
    
    \vspace{-0.4cm}
    
    \includegraphics[width=\textwidth]{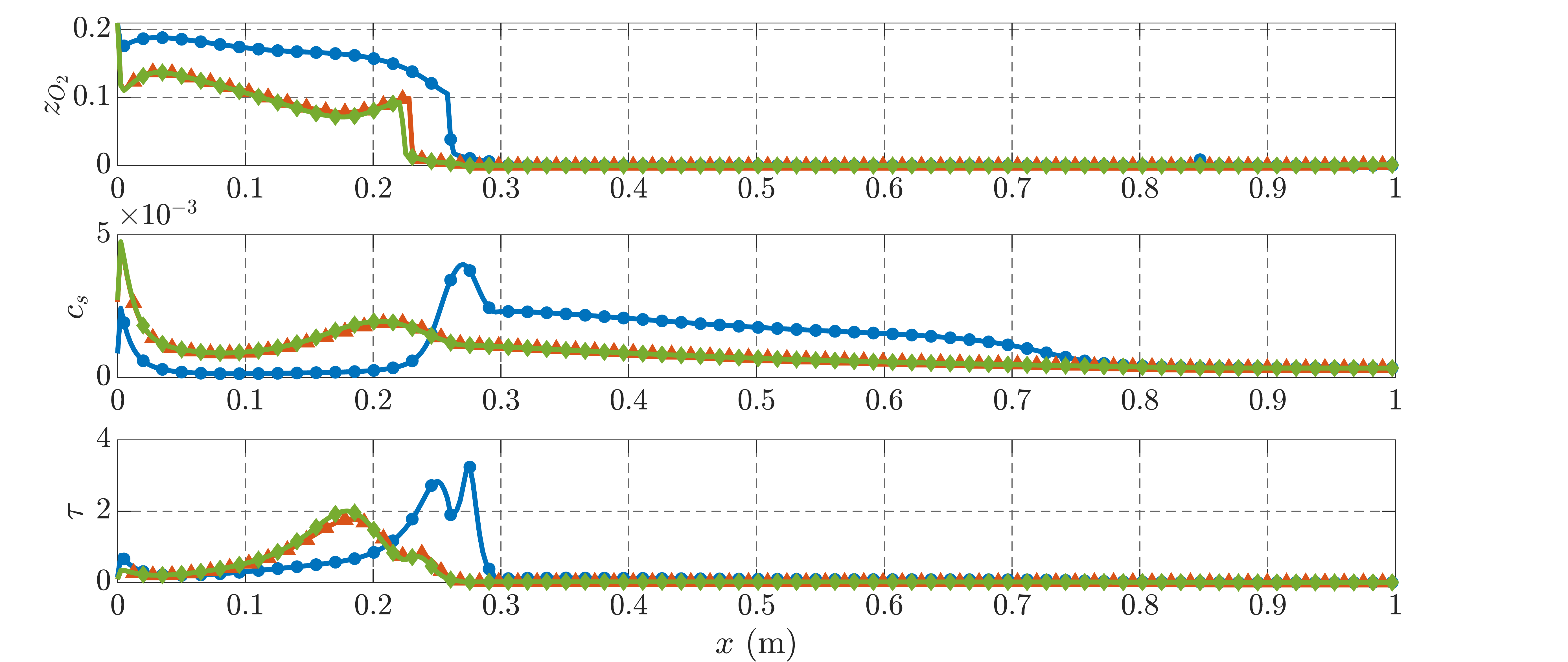}
    \caption{Hot air injection (in-situ combustion) results for Zuata oil. From top to bottom: temperature, oil saturation, water saturation, pressure, oxygen mole fraction, solid concentration and reaction rate. Using 4 components (blue circles), 8 components (orange triangles) and 52 components (green diamonds).}
    \label{fig::DeadZuata_Inj400CReac}
\end{figure*}

\begin{figure*}[t!]
    \centering
    \includegraphics[width=\textwidth]{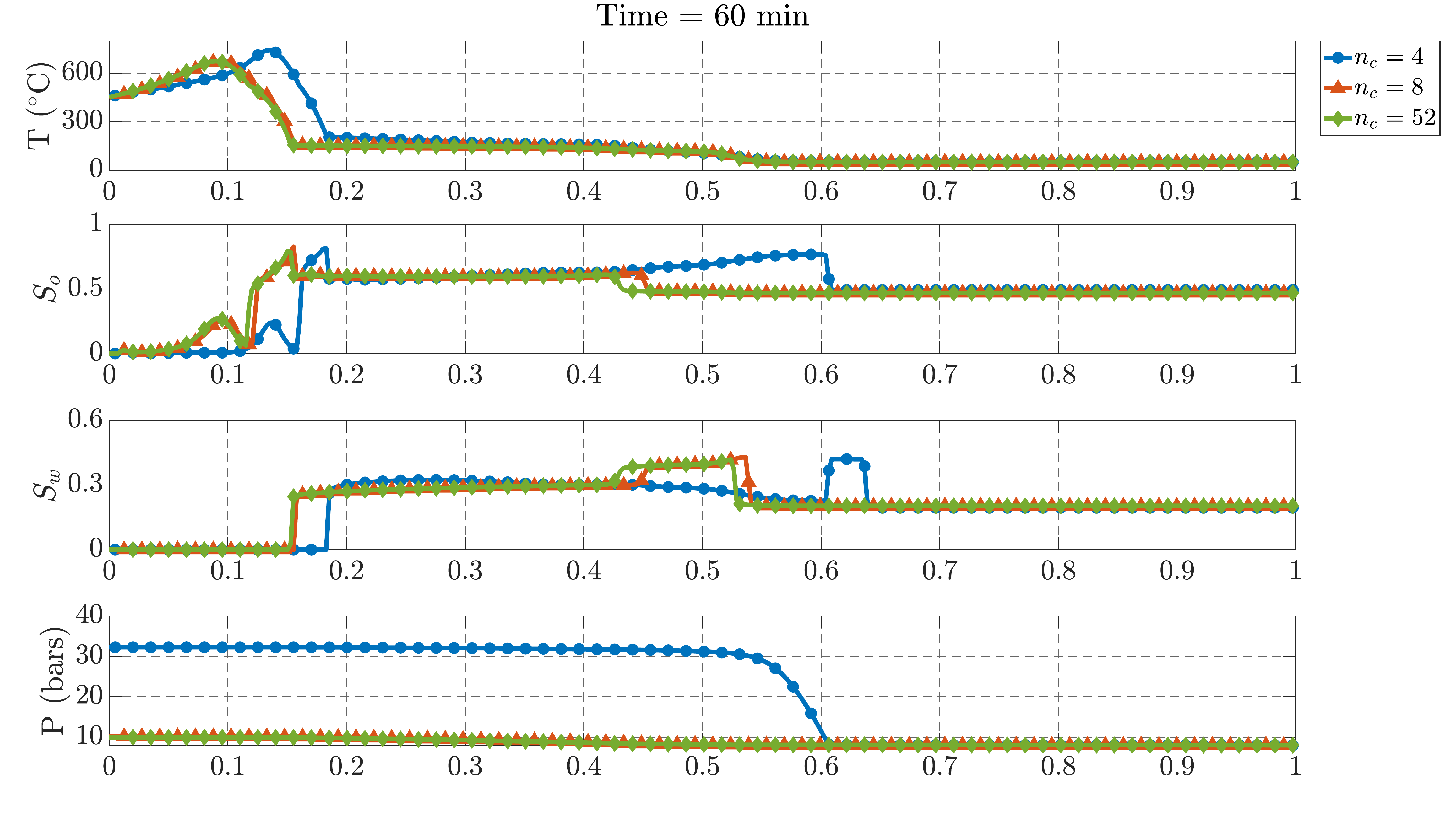}
    
    \vspace{-0.4cm}
    
    \includegraphics[width=\textwidth]{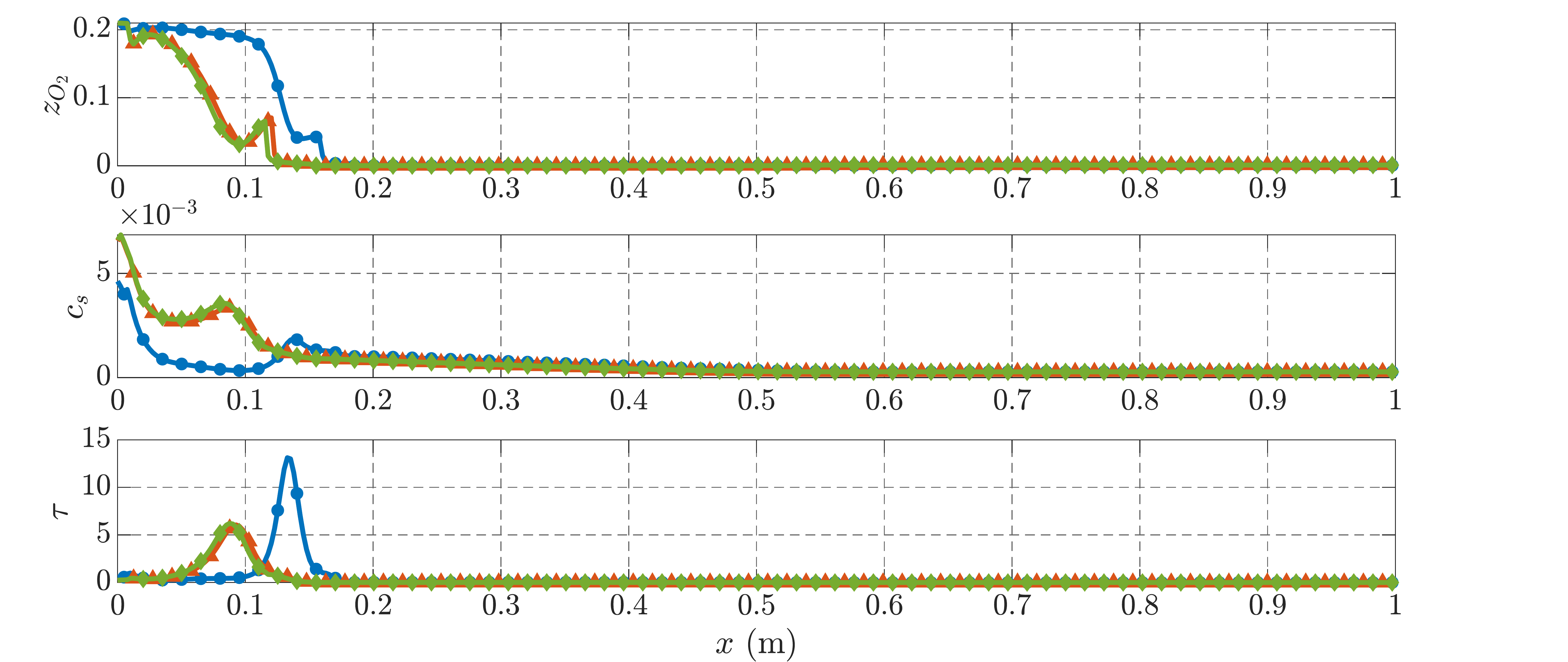}
    \caption{Hot air injection (in-situ combustion) results for MacKay oil. From top to bottom: temperature, oil saturation, water saturation, pressure, oxygen mole fraction, solid concentration and reaction rate. Using 4 components (blue circles), 8 components (orange triangles) and 52 components (green diamonds).}
    \label{fig::MacKayBitumen_Inj400CReac}
\end{figure*}

Figure~\ref{fig::DeadZuata_Inj400CReac} shows the temperature, oil saturation, water saturation, pressure, oxygen mole fraction, solid concentration and reaction rate for the Zuata oil. Figure~\ref{fig::MacKayBitumen_Inj400CReac} shows the same quantities for the MacKay oil. The same behavior is observed as for the nitrogen injection. The 4-component lumping is not capable of capturing the correct displacement behavior. The pressure is severely overestimated (6x greater), and the fronts are misaligned. The oil bank is overestimated as well, mostly due to the light oil component lumping. Finally, the reaction quantities (oxygen fraction, solid concentration and reaction rate) also do not match. 

The 8-component lumping is again performing well, for a fraction of the cost. Using the sparse direct linear solver SuperLU \citep{Li05}, the runtime for the 52-component Zuata case is slightly above 30 hours for 300 min of physical time. With 8 components, it gets down to around 1.25 hours, or 24 times faster (using a linux CentOS compute node, with Dual 8-core x64 Intel E5-2660 2.2 Ghz, 64 GB of DDR3-1600 RAM). Interestingly, the 8-components case requires a lot more Newton iterations (five time as many) than the 52-components case, showing that the lumping challenges even the numerical solvers. In this work, a small maximum timestep is prescribed to limit the numerical errors as much as possible, and the focus is on the physical phenomena. More numerical error analysis would be important, especially with the objective of allowing larger timesteps to reduce the runtime, but it lies outside the scope of this paper. \editone{The use of an iterative linear solver requires complex preconditioning strategies \citep{Li17, Roy20, Cremon20etal} and has not been considered for this study.}

\subsection{Cold Air Injection with Heaters (CAIH)}

The final case considered for lumping effects, CAIH, is another in-situ combustion problem. Combustion tube experiments typically use electric heaters to reach ignition conditions, rather than injecting hot fluids. The same model from \citet{Dechelette06} is used here, but this time inject air at the initial temperature (50$^\circ$C) and use heaters on the first 10\% of the domain during 30 minutes. The resulting heating scheme is more challenging to simulate, since it is less smooth than a continuous increase of temperature similar to the one experienced in the previous cases. It also provides more overall heat to the system, and the combustion front burns at a higher temperature and thus in a more self-sustainable manner.

\begin{figure*}[t!]
    \centering
    \includegraphics[width=\textwidth]{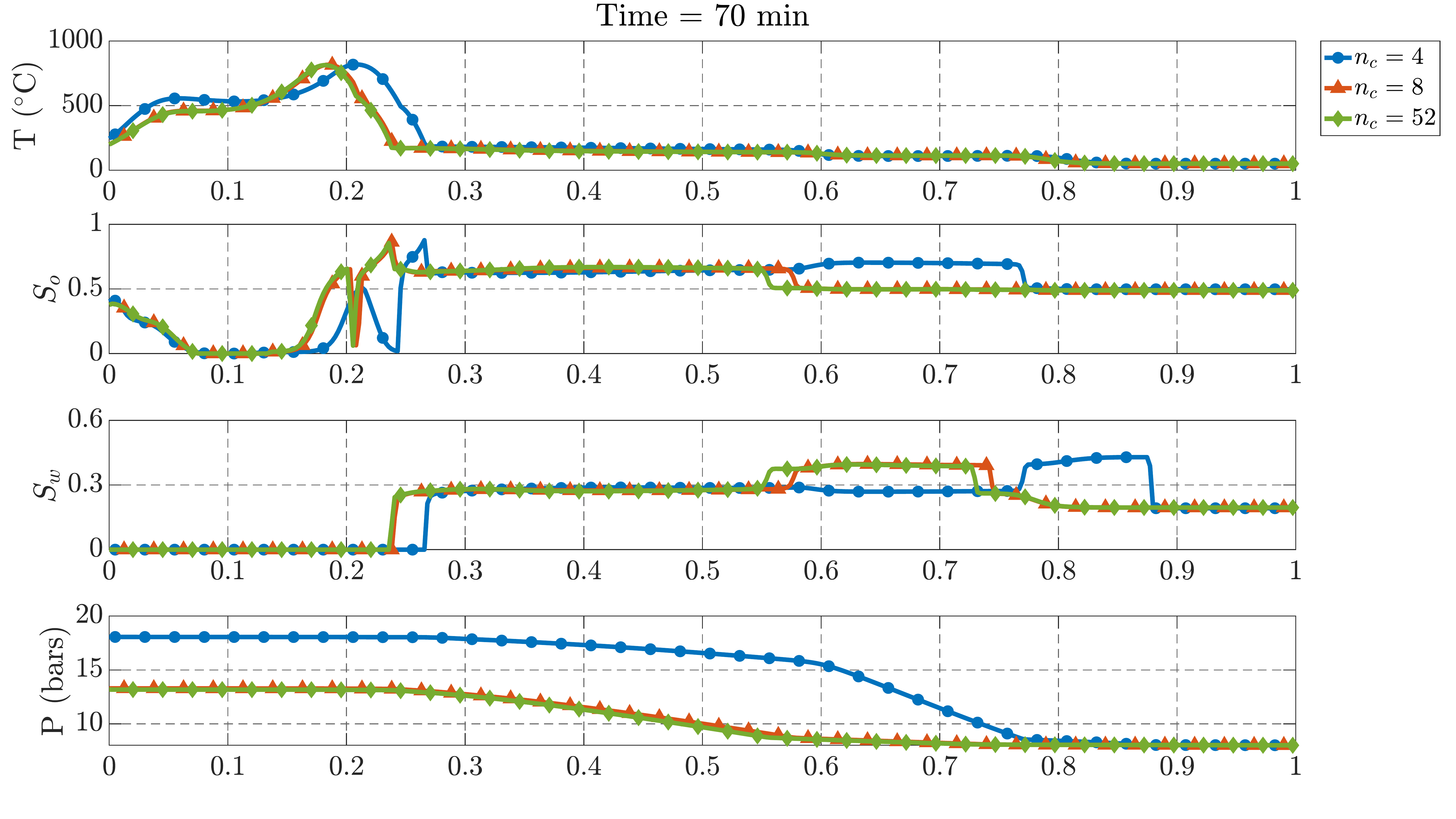}
    
    \vspace{-0.4cm}
    
    \includegraphics[width=\textwidth]{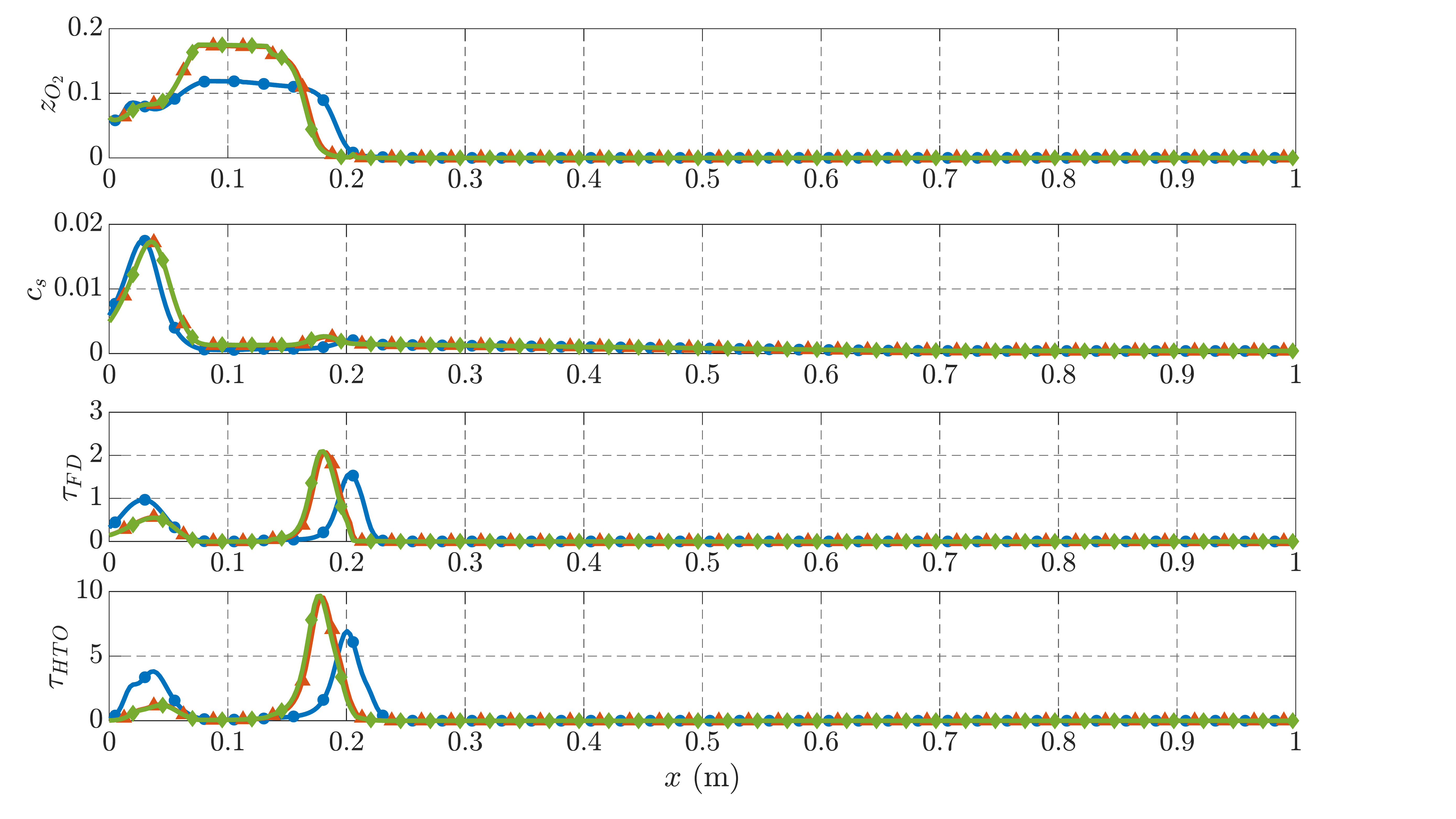}
    \caption{Cold air injection with heaters (in-situ combustion) results for Zuata oil. From top to bottom: temperature, oil saturation, water saturation, pressure, oxygen mole fraction, solid concentration and reaction rate. Using 4 components (blue circles), 8 components (orange triangles) and 52 components (green diamonds).}
    \label{fig::DeadZuata_HeatersReac}
\end{figure*}

In CAIH, the system undergoes a different heating sequence compared to simply injecting a hot fluid. A premature evaporation of heavy components is still observed with the 4-component lumping, as well as the development of a larger oil bank. Most shocks are moving downstream faster than the detailed composition results, and an overpressure appears when displacing larger liquid (oil and water) banks. Moving more oil downstream reduces the reaction rates, since the fuel deposition reaction has less fuel to consume. Figure~\ref{fig::DeadZuata_HeatersReac} shows our results for that case. 

Overall, for extra-heavy oil with a large proportion of components heavier than C$_{20}$, aggressive lumping will not allow us to capture the evaporation/condensation of the components properly.

\section{Compositional Formulations for the Reaction Scheme}
\label{sec::reactionResults}

Another way compositional effects can impact our simulations is through reactions. For ISC simulations, part of the oil is converted to solid coke during a fuel deposition reaction, which then burns during a High Temperature Oxidation (HTO) reaction \citep{Prats82}. Fuel deposition reactions are typically written in the literature with only one Oil species \citep{Dechelette06,Cinar11}, sometimes divided between light oil and heavy oil \citep{Crookston79}. When using a compositional framework, there is no straightforward mapping between the reactive components and the flow/thermodynamics components. This section studies the influence of using several compositional mappings in the first reaction of the \citet{Dechelette06} model, Eq.~\eqref{eq::FD}.

For runtime considerations and based on the results from Section~\ref{sec::lumpingResults}, the 8-components lumping is used in the remainder of this work. \editone{It should be noted that the same conclusions would be drawn for the 52-component case; this part of the work does not depend on the lumping scheme.}
Most of the classical reaction schemes have been devised using static kinetic cell experiments, which are conducted below the residual oil saturation so that the oil does not move. The absence of displacement allows the design of lumped reactions, which make the schemes more compact (around 3-6 reactions) but largely ignore phase behavior effects.
\editfour{Most of the experimental and theoretical work about heavy oil suggests that the molecular structure of the components plays an important role in the reactions \citep{Wiehe93,Belgrave93,Cinar11b}. The SARA (Saturates, Aromatics, Resins, Asphaltenes) description is typically used to characterize heavy oils \citep{Speight04,Freitag06,Kristensen09}. Those fractions represent different molecule types and exhibit different reaction behavior. To be integrated with a compositional model such as the one used here, each of the 4 SARA fractions would need to be further characterized in terms of components and given critical properties. The molecules belonging to the SARA fractions vary in size and composition, and their characterization is an open research question, preventing an easy integration into a compositional simulator. There does not seem to be a consensus on how to assign properties to those complex molecules, especially asphaltenes \citep{Powers16}. Lastly, the number of components required for all fractions would become impractical to run detailed simulations. For those reasons, those considerations were not included in this analysis.}

\editone{The numerical study conducted here does not aim to find the most suitable compositional mapping, but rather aims to illustrate that those mappings have a large influence on the results, and to inform practitioners of their importance in the search for predictive simulations.}
Our objective with this part of the work is to study the influence of modifying the mapping between reactants in the chemical reactions and thermodynamic-based components in the mass conservation problem. In the first reaction of the \citet{Dechelette06} model, a pseudo-species called Oil is the reactant. A logical compositional mapping is then to consider all components as reactants. It is well established \citep{Wiehe93,Belgrave93,Cinar11b} that not all hydrocarbons, in terms of both nature and number of carbon atoms, will actually get converted to solid fuel; only the heavier fractions of resins and asphaltenes are cracked and precipitated into fuel. This suggests that the light and medium components should not be used in this reaction if carbon number is taken into account.

\begin{figure}[htb]
    \centering
    \includegraphics[width=.5\textwidth]{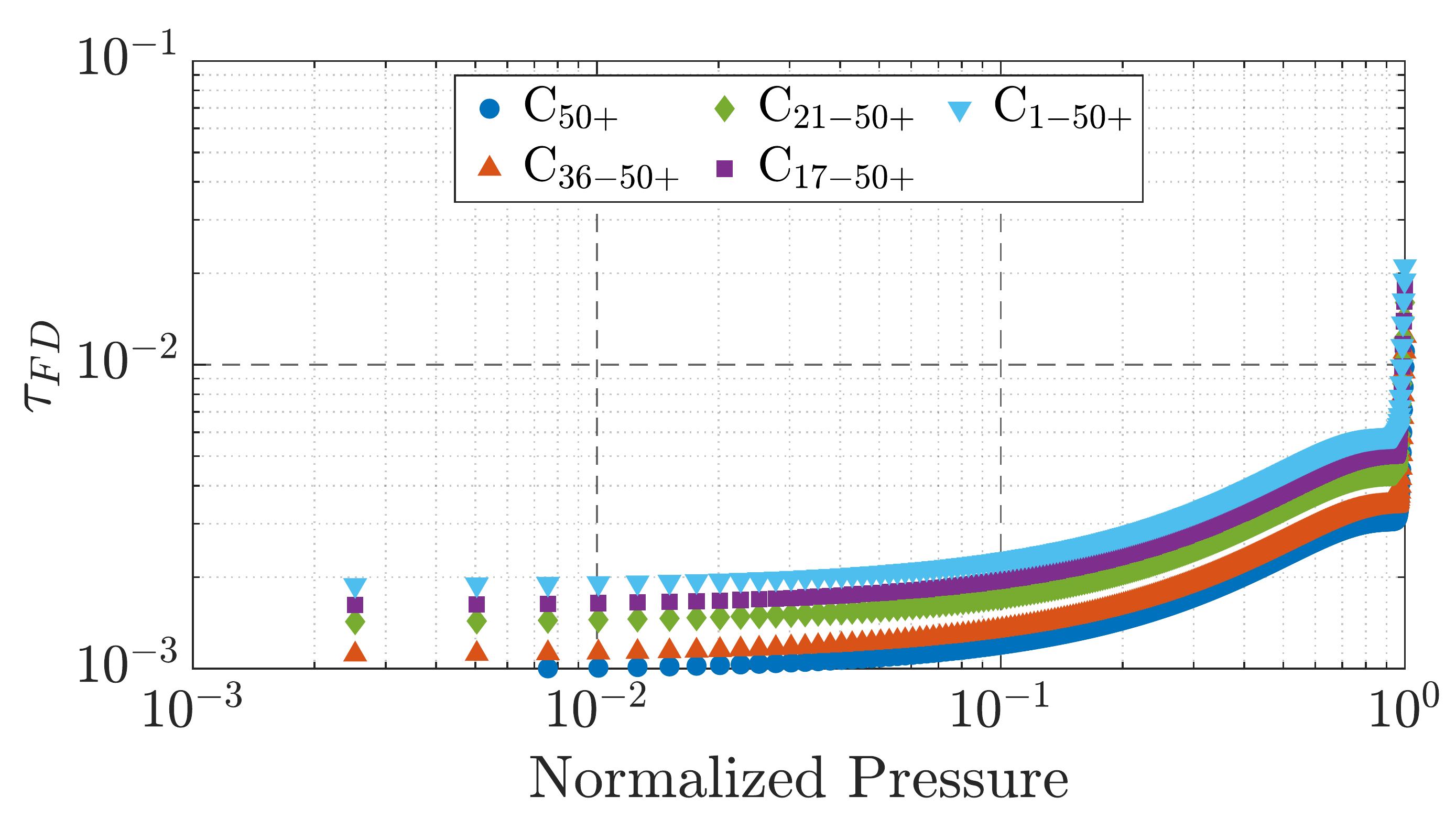}
    \includegraphics[width=.5\textwidth]{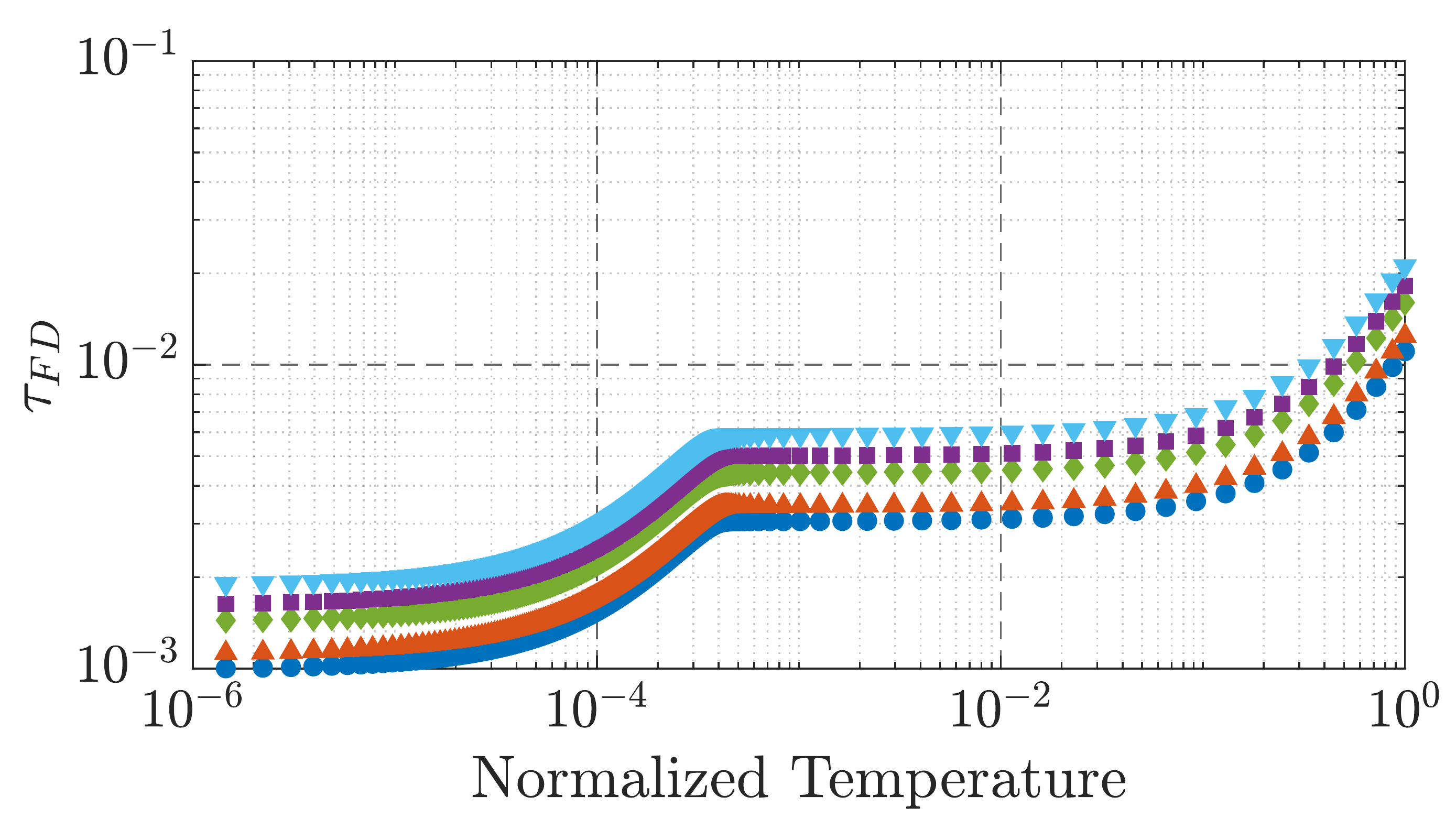}
    \caption{Reaction rate as a function of normalized pressure (top) and temperature (bottom). Using C$_{50+}$ (blue circles), C$_{36-50+}$ (orange upward-pointing triangles), C$_{22-50+}$ (green diamonds), C$_{17-50+}$ (purple squares) and C$_{2-50+}$ (cyan downward-pointing triangles) in the fuel deposition reaction.}
    \label{fig::reactionRate}
\end{figure}

The cases considered are presented in Table~\ref{tab::reacCases}.
\begin{table}[b!]
    \centering
    \caption{Reactive components for different cases}
    \begin{tabular}{lcc}
        \toprule
        Case & Reactive Components & Inert Components \\
        \midrule
        1 & C$_{50+}$ & C$_{1-49}$ \\
        2 & C$_{36-50+}$ & C$_{1-35}$ \\
        3 & C$_{22-50+}$ & C$_{1-21}$ \\
        4 & C$_{17-50+}$ & C$_{1-16}$ \\
        5 & C$_{2-50+}$ & C$_{1}$ \\
        \bottomrule
    \end{tabular}
    \label{tab::reacCases}
\end{table}
Note that for the extra-heavy oils in this work, methane is not present. Following guidelines in \citet{Kovscek13}, at most 8\% by mass of the oil is allowed to react. The fact that that fraction is calculated by mass has severe implications on the reactions, due to molecular weight effects. The temperature dependent part of the reaction rate, or rate constant, is computed as
\begin{equation}
    k(T) = A\exp\left(\dfrac{-E_a}{RT}\right),
\end{equation}
with $A$ the frequency factor, $E_a$ the activation energy and $R$ the ideal gas constant. Those quantities only depend on the reaction, so for a given temperature the rate constant is the same regardless of the number of reacting components. The reaction rate is computed as
\begin{equation}
    \tau = A\exp\left(\dfrac{-E_a}{RT}\right)c_XP^{O_2},
\end{equation}
with $c_X$ the molar concentration of the reactant, and $P^{O_2}$ the partial pressure of oxygen. Since the allocation of the portion of oil allowed to react is done by mass, the number of moles of reactants is different for each case, and due to molecular weight effects it grows with the number of reacting components. The case with only one reactant (C$_{50+}$) has almost twice fewer moles of reactants (92\% less) than the case with all components. Since the reaction rate depends on the concentration, less fuel is deposited even if the temperature and pressure are equal. Figure~\ref{fig::reactionRate} shows the reaction rate as a function of normalized pressure (a) and temperature (b) for an early timestep (2 minutes), confirming the systematic increase in reaction rate when more components are allowed to react.

Results from two test cases are shown, identical to the Hot Air Injection and Cold Air Injection with Heaters problems from the previous section, using Zuata oil. 
\begin{figure*}[t!]
    \centering
    \includegraphics[width=\textwidth]{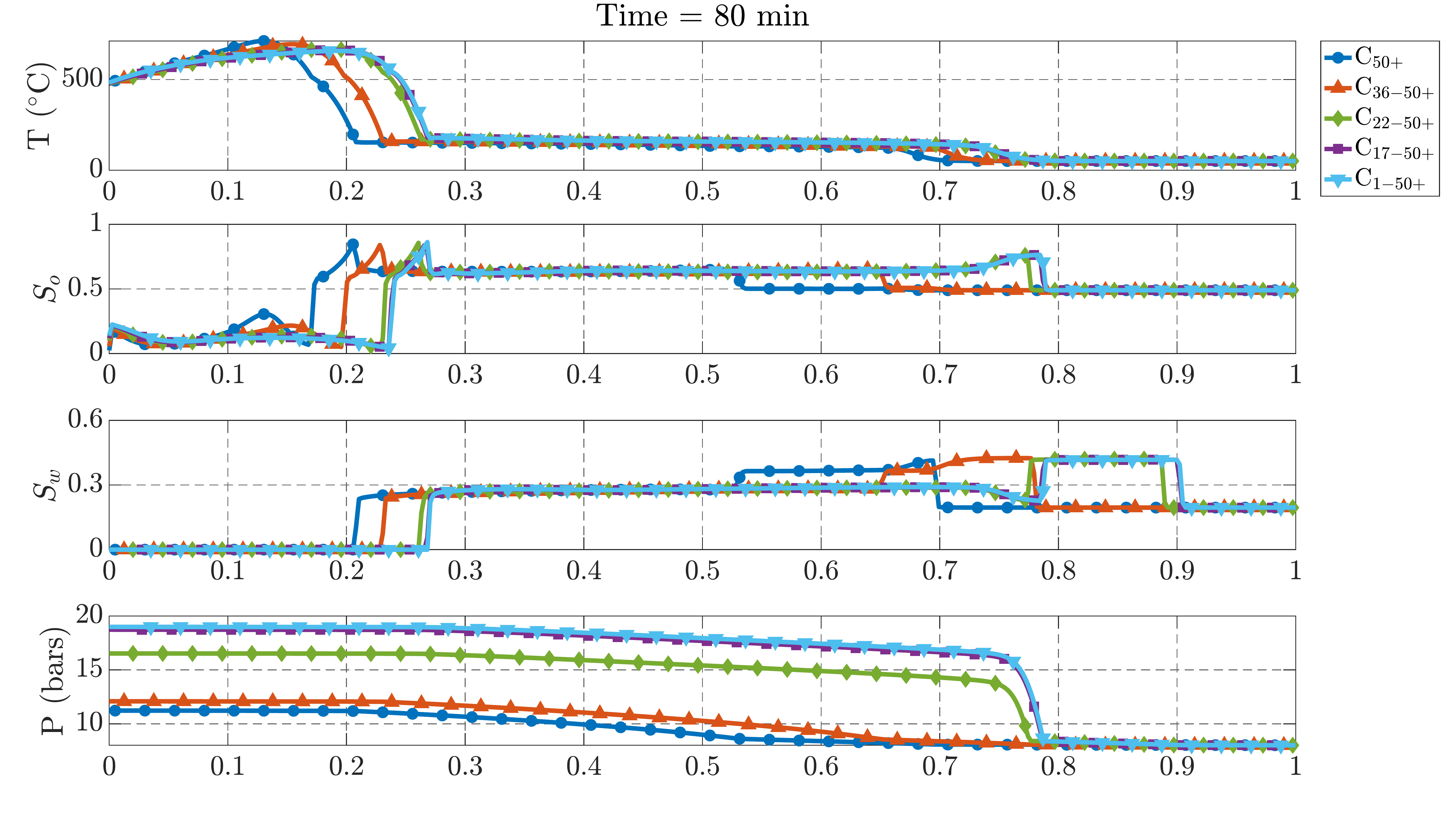}
    
    \vspace{-0.4cm}
    
    \includegraphics[width=\textwidth]{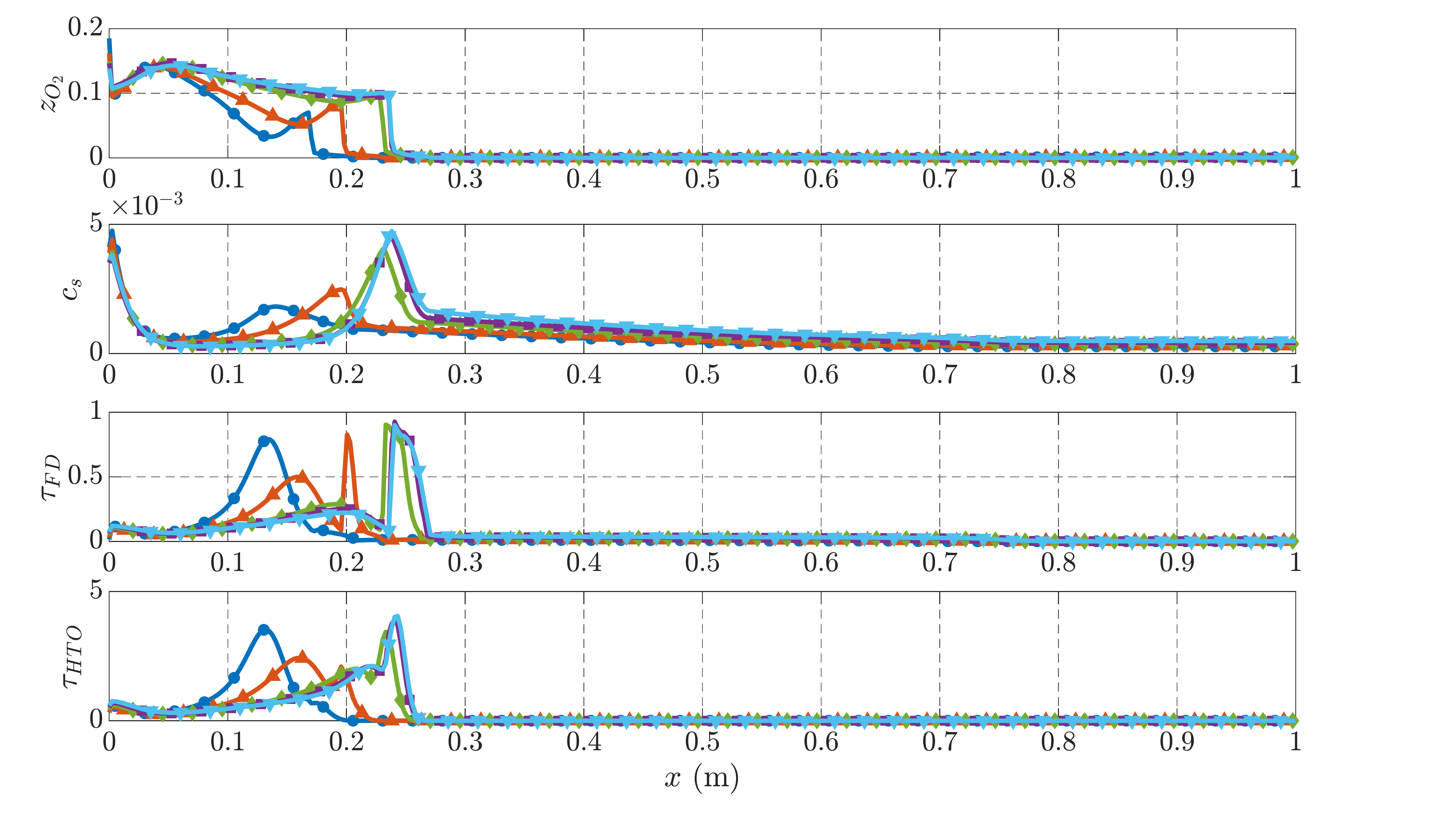}
    \caption{Hot air injection results for Zuata oil and different components reacting. From top to bottom: temperature, oil saturation, water saturation, pressure, oxygen mole fraction, solid concentration, reaction rate for the fuel deposition reaction and reaction rate for the HTO reactions. Using C$_{50+}$ (blue circles), C$_{36-50+}$ (orange upward-pointing triangles), C$_{22-50+}$ (green diamonds), C$_{17-50+}$ (purple squares) and C$_{2-50+}$ (cyan downward-pointing triangles) in the fuel deposition reaction.}
    \label{fig::wells1}
\end{figure*}
\begin{figure*}[t!]
    \centering
    \includegraphics[width=\textwidth]{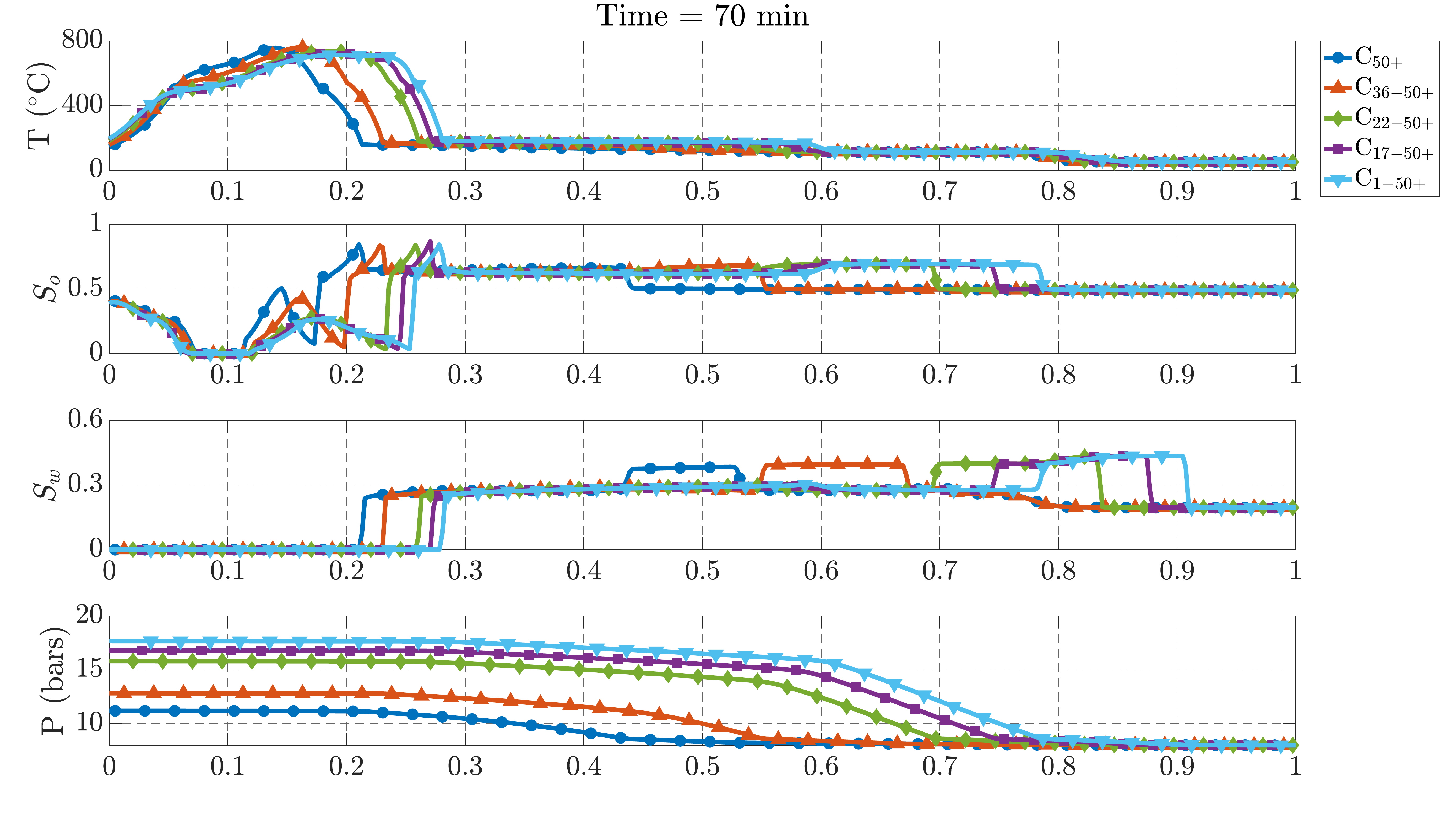}
    
    \vspace{-0.4cm}
    
    \includegraphics[width=\textwidth]{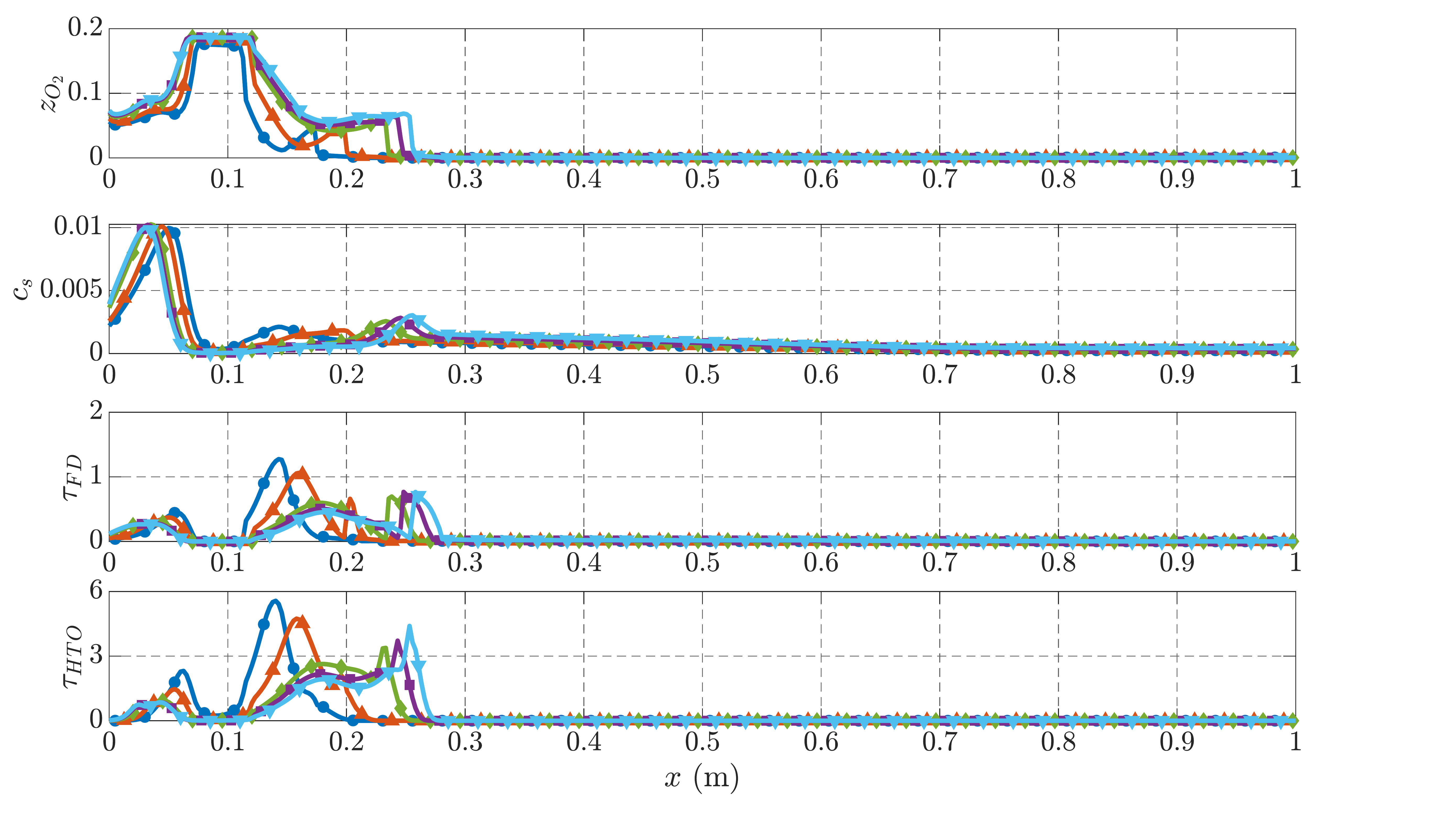}
    \caption{Cold air injection with heaters results for Zuata oil and different components reacting. From top to bottom: temperature, oil saturation, water saturation, pressure, oxygen mole fraction, solid concentration, reaction rate for the fuel deposition reaction and reaction rate for the HTO reactions. Using C$_{50+}$ (blue circles), C$_{36-50+}$ (orange upward-pointing triangles), C$_{22-50+}$ (green diamonds), C$_{17-50+}$ (purple squares) and C$_{2-50+}$ (cyan downward-pointing triangles) in the fuel deposition reaction.}
    \label{fig::heaters1}
\end{figure*}
Figure~\ref{fig::wells1} shows our results from the hot air injection test case previously described. The interplay between reactions and evaporation leads to an increased speed for the reaction front (both fuel deposition and HTO) as well as an increased oil bank size. Due to the larger liquid displacement, the system pressurizes and reaches pressures that would not be possible to achieve in the lab (in excess of 15 bars or 200 psi). The fuel deposition reaction shows a wider profile due to the sequential evaporation of components that can now react. Figure~\ref{fig::heaters1} shows our results for the CAIH case, for which the same behavior is observed: faster fronts, larger oil banks and larger pressure. Since the system of equations is highly non-linear and tightly coupled, it is difficult to speculate on which physical process is the most influential. For the cases with more reactants, some of them are now evaporated and will move according to their $K$-value front speed. The combination of these effects with the increased reaction rate affects the reaction front propagation speed.

This section of the work attempts to understand the complex interaction between thermodynamic-based components used in the mass conservation and phase behavior problem and the reactants appearing in reaction schemes devised under static conditions. It is illustrated that there is a significant impact on the solution of the non-linear, coupled PDEs when the number and type of reactants are changed.

\section{Conclusions \& Future Work}
\label{sec::conclusion}


\edit{In this work, two compositional effects arising when using a fully compositional oil description for thermal and reactive simulations are investigated. A new fully compositional, thermal and reactive simulator is used to run detailed compositions involving more than 50 hydrocarbons, which are used as reference solutions for those complex dynamic cases.}

\editone{The first section studies the interplay between a fully compositional model and strong thermal effects, especially in the context of lumping.} First, the influence of the lumping procedure on 1D thermal recovery processes at laboratory scale is demonstrated. \edit{Multiple lumping schemes tuned to match the properties of the initial oils are considered in this work. Although the 2-component schemes did not yield very good matches, both the 4- and 8-component schemes gave virtually the same results as the full 52-component case to compute the phase diagram of the initial oil. When running the dynamic 1D problem, the 4-component lumping led to large inaccuracies, due to its inability to correctly approximate the phase behavior of the much heavier oils encountered during the simulation involving large thermal effects. The 8-component lumping did lead to accurate results, for a fraction of the cost of the detailed simulations. This works shows for the first time that matching the properties of the initial oil is a necessary but not sufficient condition to accurately capture dynamic processes involving thermal and reactive changes. The same behavior was observed with and without oxidation reactions and heaters, using two different extra-heavy oils.}

\editone{The second section focused on the relationship between components from the compositional model, and species in the reaction schemes.} It is showed that different compositional mappings lead to different front speeds and oil banks sizes. 8\% of the mass is allowed to react, leading to increased reaction rates when ligther components are reacting. The system also pressurizes more when medium and light components are included in the reactions.

An important discussion is that this study has been conducted at laboratory scale, and its conclusion are only applicable to combustion tube experiments. The upscaling process from laboratory to field scale is an open research question, due to both numerical (temperature averaging from finite volume coupled with Arrhenius kinetics is problematic \citep{Nissen15}) and physical phenomena (dimensionless numbers such as P\'eclet and Damk\"ohler cannot be easily matched when the characteristic dimension changes). \editfour{The observations made through this work would need to be re-tested on field cases, which also exhibit a much larger pressure, leading to a possibly different phase behavior.}

For future work, considering the impact of more complicated reaction schemes would open more research avenues. The \citet{Dechelette06} model only considers one hybrid fuel deposition and low temperature oxidation reaction involving oil. Other models from \citet{Crookston79} or \citet{Cinar11b} could provide more interesting interactions with phase behavior models. Running cases in 2- or 3-dimensions could also yield interesting insights, but may require a switch to a preconditioned iterative linear solver for runtime considerations \citep{Li17,Cremon20etal,Roy20}. Including heterogeneity would also be interesting.


\section{Acknowledgements}

The authors wish to thank Prof. Anthony R. Kovscek (Stanford), Prof. Hamdi A. Tchelepi (Stanford) and Prof. Dan V. Nichita (Universit\'e de Pau et des Pays de l'Adour) for numerous ideas and fruitful discussions, and Ecopetrol for financial support.


\bibliographystyle{spbasic}
\bibliography{bib_final}


\end{document}